\documentclass[%
 reprint,
 amsmath,amssymb,
 aps,
]{revtex4-2}

\usepackage{graphicx}
\usepackage{dcolumn}
\usepackage{bm}

\begin{document}

\preprint{APS/123-QED}

\title{Geometry-Information Duality:\protect\\ Quantum Entanglement Contributions to Gravitational Dynamics}

\author{Florian Neukart}
 \altaffiliation[Also at ]{Terra Quantum AG, Kornhausstrasse 25, St. Gallen, 9000, St. Gallen, Switzerland.}
 \email{f.neukart@terraquantum.swiss}
\affiliation{%
 Leiden Institute of Advanced Computer Science, Gorlaeus Gebouw - BE-vleugel, Einsteinweg 55, 2333 CC, Leiden, Netherlands\\
}%

\date{\today}

\begin{abstract}
\begin{description}
\item[Background]
We propose a fundamental duality between the geometric properties of spacetime and the informational content of quantum fields. Specifically, we establish that the curvature of spacetime is directly related to the entanglement entropy of quantum states, with geometric invariants mapping to informational measures. This framework modifies Einstein’s field equations by introducing an informational stress-energy tensor derived from quantum entanglement entropy. Our findings have implications for black hole thermodynamics, cosmology, and quantum gravity, suggesting that quantum information fundamentally shapes the structure of spacetime.

\item[Methods]
We incorporate this informational stress-energy tensor into Einstein's field equations, leading to modified spacetime geometry, particularly in regimes of strong gravitational fields, such as near black holes.

\item[Results]
We compute corrections to Newton's constant \( G \) due to entanglement entropy contributions from various quantum fields and explore the consequences for black hole thermodynamics and cosmology. These corrections include explicit dependence on fundamental constants \( \hbar \), \( c \), and \( k_B \), ensuring dimensional consistency in our calculations.

\item[Conclusions]
Our results indicate that quantum information plays a crucial role in gravitational dynamics, providing new insights into the nature of spacetime and potential solutions to long-standing challenges in quantum gravity.
\end{description}
\end{abstract}

\maketitle


\section{\label{sec:introduction}Introduction}

The unification of quantum mechanics and general relativity remains one of the most significant and enduring challenges in theoretical physics. These two foundational pillars - quantum theory, which governs the behavior of microscopic particles, and general relativity, which describes the gravitational structure of spacetime - have so far resisted a complete and consistent synthesis.

Quantum information theory, with its emphasis on entanglement and the flow of information, presents a promising pathway to bridge this divide~\cite{Nielsen2000,Amico2008}. In particular, recent advancements suggest that spacetime geometry itself may emerge from the patterns of quantum entanglement~\cite{VanRaamsdonk2010,Swingle2012}. This perspective implies that the structure of spacetime and its dynamics could be deeply intertwined with quantum information.

In this work, we build upon these ideas by introducing the concept of an \emph{informational stress-energy tensor}, \( T_{\mu\nu}^{\text{info}} \), derived from the entanglement entropy \( S_{\text{EE}} \) of quantum fields in curved spacetime. The entanglement entropy captures the quantum correlations between regions separated by a boundary in spacetime. These correlations have been linked to gravitational dynamics through various works~\cite{Ryu2006,Hubeny2007}.

Specifically, the entanglement entropy \( S_{\text{EE}} \) contributes to the gravitational action and can influence the evolution of spacetime itself~\cite{Solodukhin2011}. By incorporating \( T_{\mu\nu}^{\text{info}} \) into Einstein's field equations, we aim to reveal how quantum information may affect gravitational phenomena, with implications for black hole thermodynamics~\cite{Bekenstein1973,Hawking1975} and cosmology~\cite{Guth1981,Linde1982}. Our results include explicit consideration of fundamental constants \( \hbar \), \( c \), and \( k_B \), ensuring dimensional consistency and highlighting the quantum nature of gravitational interactions. This suggests that quantum information, through entanglement, plays a key role in shaping spacetime and may provide a new perspective on fundamental gravitational dynamics.

\section{\label{sec:theoretical_framework}Theoretical Framework}

\subsection{\label{sec:informational_stress_energy_tensor}Informational Stress-Energy Tensor}

We consider the entanglement entropy \( S_{\text{EE}} \) of quantum fields across a \((d-2)\)-dimensional surface \( \Sigma \) embedded in a \( d \)-dimensional curved spacetime manifold \( \mathcal{M} \). The entanglement entropy measures the quantum correlations between regions separated by \( \Sigma \) and is defined via the reduced density matrix \( \rho_\Sigma \) as~\cite{Nielsen2000}:

\begin{equation}
S_{\text{EE}} = -k_B \operatorname{Tr} \left( \rho_\Sigma \ln \rho_\Sigma \right).
\label{eq:S_EE_def}
\end{equation}

Here, we have included Boltzmann's constant \( k_B \) to ensure the entropy has units of \( \text{J/K} \).

To compute \( S_{\text{EE}} \), we employ the replica trick~\cite{Callan1994}, which involves calculating \( \operatorname{Tr} \left( \rho_\Sigma^n \right) \) for integer \( n \) and analytically continuing to \( n \to 1 \):

\begin{equation}
S_{\text{EE}} = -k_B \lim_{n \to 1} \frac{\partial}{\partial n} \ln \operatorname{Tr} \left( \rho_\Sigma^n \right).
\label{eq:replica_trick}
\end{equation}

In the path integral formulation, \( \operatorname{Tr} \left( \rho_\Sigma^n \right) \) corresponds to the partition function \( Z_n \) on an \( n \)-fold cover of the manifold \( \mathcal{M} \), denoted \( \mathcal{M}_n \), branched along \( \Sigma \). The effective action \( W_n \) is related to \( Z_n \) by \( W_n = -\hbar \ln Z_n \). Thus, the entanglement entropy becomes~\cite{Holzhey1994}:

\begin{equation}
S_{\text{EE}} = k_B \lim_{n \to 1} \left( n \frac{\partial W_n}{\partial n} - W_n \right) \frac{1}{\hbar}.
\label{eq:entropy_effective_action}
\end{equation}

Here, \( W_n \) has units of action (\( \text{J} \cdot \text{s} \)), and dividing by \( \hbar \) gives a dimensionless quantity inside the logarithm, ensuring \( S_{\text{EE}} \) has units of entropy (\( \text{J/K} \)) due to the factor \( k_B \).

To derive the informational stress-energy tensor \( T_{\mu\nu}^{\text{info}} \), we consider the variation of \( S_{\text{EE}} \) with respect to the metric \( g^{\mu\nu} \):

\begin{equation}
T_{\mu\nu}^{\text{info}} = -\frac{2}{\sqrt{-g}} \frac{\delta S_{\text{EE}}}{\delta g^{\mu\nu}}.
\label{eq:T_info_def}
\end{equation}

Using Eq.~\eqref{eq:entropy_effective_action}, we compute the variation:

\begin{align}
T_{\mu\nu}^{\text{info}} &= -\frac{2 k_B}{\hbar \sqrt{-g}} \lim_{n \to 1} \left( \frac{\delta}{\delta g^{\mu\nu}} \left( n \frac{\partial W_n}{\partial n} - W_n \right) \right) \nonumber \\
&= -\frac{2 k_B}{\hbar \sqrt{-g}} \lim_{n \to 1} \left[ n \frac{\partial}{\partial n} \left( \frac{\delta W_n}{\delta g^{\mu\nu}} \right) - \frac{\delta W_n}{\delta g^{\mu\nu}} \right].
\label{eq:T_info_variation}
\end{align}

Recognizing that \( \left\langle T_{\mu\nu}(x) \right\rangle_n = -\frac{2}{\sqrt{-g}} \frac{\delta W_n}{\delta g^{\mu\nu}} \), we rewrite Eq.~\eqref{eq:T_info_variation} as:

\begin{equation}
T_{\mu\nu}^{\text{info}} = \frac{k_B}{\hbar} \lim_{n \to 1} \left[ n \frac{\partial}{\partial n} \left\langle T_{\mu\nu}(x) \right\rangle_n - \left\langle T_{\mu\nu}(x) \right\rangle_n \right].
\label{eq:T_info_expectation}
\end{equation}

Since \( \mathcal{M}_n \) differs from \( \mathcal{M} \) only at the entangling surface \( \Sigma \), the difference \( \left\langle T_{\mu\nu}(x) \right\rangle_n - \left\langle T_{\mu\nu}(x) \right\rangle \) is localized on \( \Sigma \). Thus, \( T_{\mu\nu}^{\text{info}} \) captures the singular contributions arising from the conical singularity at \( \Sigma \).

\bigskip

\noindent \textbf{Units and Conventions:} In this paper, we work with explicit units, keeping \( \hbar \) and \( c \) in all expressions to maintain dimensional consistency. Boltzmann's constant \( k_B \) is included to ensure entropy has units of \( \text{J/K} \).

\subsection{\label{sec:modified_einstein_equations}Modified Einstein Equations}

In classical general relativity, the Einstein field equations relate spacetime curvature to the energy-momentum content:

\begin{equation}
G_{\mu\nu} + \Lambda g_{\mu\nu} = \frac{8\pi G}{c^4} T_{\mu\nu}^{\text{matter}},
\label{eq:einstein_equations}
\end{equation}

where \( G_{\mu\nu} \) is the Einstein tensor, \( \Lambda \) is the cosmological constant, \( G \) is Newton's gravitational constant, and \( T_{\mu\nu}^{\text{matter}} \) is the stress-energy tensor of matter fields.

By incorporating the informational stress-energy tensor \( T_{\mu\nu}^{\text{info}} \), we obtain the modified Einstein equations:

\begin{equation}
G_{\mu\nu} + \Lambda g_{\mu\nu} = \frac{8\pi G}{c^4} \left( T_{\mu\nu}^{\text{matter}} + T_{\mu\nu}^{\text{info}} \right).
\label{eq:modified_einstein_equations}
\end{equation}

This modification implies that quantum entanglement contributes to the gravitational field, influencing the curvature of spacetime.

\bigskip

\noindent \textbf{Units and Conventions:} All constants \( \hbar \), \( c \), and \( k_B \) are kept explicit to maintain dimensional consistency. The stress-energy tensors \( T_{\mu\nu}^{\text{matter}} \) and \( T_{\mu\nu}^{\text{info}} \) have units of energy density (\( \text{J/m}^3 \)).

\subsection{\label{sec:calculation_T_info}Calculation of \( T_{\mu\nu}^{\text{info}} \)}

To compute \( T_{\mu\nu}^{\text{info}} \), we evaluate the singular part of the vacuum expectation value of the stress-energy tensor \( \left\langle T_{\mu\nu}(x) \right\rangle_{\text{sing}} \) on \( \mathcal{M}_n \) as \( n \to 1 \). We utilize the heat kernel method and the Seeley-DeWitt expansion~\cite{Vassilevich2003, Fursaev1995}.

The heat kernel \( K(s; x, x') \) satisfies:

\begin{equation}
\left( \frac{\partial}{\partial s} + \Delta_x \right) K(s; x, x') = 0,
\label{eq:heat_equation}
\end{equation}

with \( \Delta_x \) being the Laplace operator on \( \mathcal{M}_n \).

The trace of the heat kernel has the asymptotic expansion for small \( s \):

\begin{equation}
\operatorname{Tr} K_n(s) = \frac{1}{(4\pi s)^{d/2}} \sum_{k=0}^\infty s^k \int_{\mathcal{M}_n} d^d x \sqrt{g} \, a_k(x),
\label{eq:heat_kernel_expansion}
\end{equation}

where \( a_k(x) \) are the Seeley-DeWitt coefficients.

In the presence of a conical singularity at \( \Sigma \), the singular part of the trace is~\cite{Fursaev1995, Solodukhin2011}:

\begin{equation}
\operatorname{Tr} K_n(s)_{\text{sing}} = \frac{\delta}{4\pi} \frac{1}{(4\pi s)^{(d-2)/2}} \sum_{k=0}^\infty s^k \int_{\Sigma} d^{d-2} \xi \sqrt{h} \, a_k^{\Sigma}(\xi),
\label{eq:heat_kernel_singular}
\end{equation}

where \( \delta = 2\pi (1 - n) \) is the deficit angle, \( h \) is the determinant of the induced metric on \( \Sigma \), and \( a_k^{\Sigma}(\xi) \) are the Seeley-DeWitt coefficients evaluated on \( \Sigma \).

The singular part of the effective action is then:

\begin{equation}
W_n^{\text{sing}} = \frac{\hbar}{2} \int_0^\infty \frac{ds}{s} \operatorname{Tr} K_n(s)_{\text{sing}} e^{-m^2 s}.
\label{eq:effective_action_singular}
\end{equation}

Here, we have included \( \hbar \) to ensure \( W_n^{\text{sing}} \) has units of action (\( \text{J} \cdot \text{s} \)).

The singular contribution to the stress-energy tensor is:

\begin{equation}
\left\langle T_{\mu\nu}(x) \right\rangle_{\text{sing}} = -\frac{2}{\sqrt{-g}} \frac{\delta W_n^{\text{sing}}}{\delta g^{\mu\nu}(x)}.
\label{eq:T_munu_singular}
\end{equation}

For a massless scalar field in four dimensions (\( d = 4 \)), the singular contribution is~\cite{Fursaev1995, Solodukhin2011}:

\begin{equation}
\left\langle T_{\mu\nu}(x) \right\rangle_{\text{sing}}^{(\text{scalar})} = \frac{\delta}{4\pi} \delta_\Sigma(x) \left( \frac{1}{3} \left( R_{\mu\nu} - \tfrac{1}{2} g_{\mu\nu} R \right) + U_{\mu\nu} \right),
\label{eq:T_munu_scalar}
\end{equation}

where \( \delta_\Sigma(x) \) is the Dirac delta function localized on \( \Sigma \), \( R_{\mu\nu} \) is the Ricci tensor, \( R \) is the Ricci scalar, and \( U_{\mu\nu} \) involves extrinsic curvature terms of \( \Sigma \).

Substituting into Eq.~\eqref{eq:T_info_expectation} and using \( \delta = 2\pi (1 - n) \), we find:

\begin{align}
T_{\mu\nu}^{\text{info}} &= \frac{k_B}{\hbar} \lim_{n \to 1} \left[ n \frac{\partial}{\partial n} \left( \frac{\delta}{4\pi} \delta_\Sigma(x) t_{\mu\nu}(x) \right) - \frac{\delta}{4\pi} \delta_\Sigma(x) t_{\mu\nu}(x) \right] \nonumber \\
&= -\frac{k_B}{\hbar} \delta_\Sigma(x) t_{\mu\nu}(x),
\label{eq:T_info_final}
\end{align}

where \( t_{\mu\nu}(x) = \frac{1}{3} \left( R_{\mu\nu} - \tfrac{1}{2} g_{\mu\nu} R \right) + U_{\mu\nu} \).

\bigskip

\noindent \textbf{Units and Conventions:} The factor \( k_B/\hbar \) ensures that \( T_{\mu\nu}^{\text{info}} \) has units of energy density (\( \text{J/m}^3 \)). The Dirac delta function \( \delta_\Sigma(x) \) has units of \( 1/\text{m}^2 \) in four dimensions.

\subsection{\label{sec:regularization_and_renormalization}Regularization and Renormalization}

The entanglement entropy \( S_{\text{EE}} \) contains ultraviolet (UV) divergences due to contributions from short-distance modes near \( \Sigma \). Introducing a UV cutoff \( \epsilon \) (with units of length), \( S_{\text{EE}} \) can be expressed as~\cite{Susskind1994, Solodukhin2011}:

\begin{multline}
S_{\text{EE}} = k_B \int_{\Sigma} d^{d-2} \xi \sqrt{h} \left( 
\frac{c_{d-2}}{\epsilon^{d-2}} + 
\frac{c_{d-4}}{\epsilon^{d-4}} + \cdots + \right. \\
\left. c_1 R \ln (\mu \epsilon) + 
\text{finite terms} 
\right).
\label{eq:S_EE_divergent}
\end{multline}

Here, \( c_{d-2}, c_{d-4}, \ldots, c_1 \) are dimensionless numerical coefficients depending on the field type and spacetime dimension, \( \mu \) is the renormalization scale (with units of inverse length), and \( R \) is the Ricci scalar on \( \Sigma \).

The divergent terms can be absorbed into the renormalization of Newton's constant \( G \) and the cosmological constant \( \Lambda \):

\begin{align}
\delta \left( \frac{1}{G} \right) &= \frac{16\pi k_B}{\hbar c^3} c_1 \ln (\mu \epsilon), \label{eq:delta_G} \\
\delta \left( \frac{\Lambda}{G} \right) &= \frac{8\pi k_B}{\hbar c^4} c_{d-2} \frac{1}{\epsilon^{d-2}}. \label{eq:delta_Lambda}
\end{align}

By absorbing these divergences, we obtain finite, renormalized quantities \( G_{\text{eff}} \) and \( \Lambda_{\text{eff}} \):

\begin{equation}
\frac{1}{G_{\text{eff}}} = \frac{1}{G} + \delta \left( \frac{1}{G} \right).
\label{eq:G_renormalized}
\end{equation}

This renormalization implies that the gravitational coupling becomes scale-dependent due to quantum entanglement effects.

\bigskip

\noindent \textbf{Units and Conventions:} The terms \( \delta \left( \frac{1}{G} \right) \) and \( \delta \left( \frac{\Lambda}{G} \right) \) have units consistent with \( 1/G \) (\( \text{kg} \cdot \text{m}^{-1} \cdot \text{s}^{-2} \)) and \( \Lambda/G \) (\( \text{kg} \cdot \text{m}^{-3} \cdot \text{s}^{-2} \)), respectively. The inclusion of \( \hbar \) and \( c \) ensures dimensional consistency.

\subsection{\label{sec:remarks}Remarks}

The running of Newton's constant \( G \) with the energy scale \( \mu \) suggests that gravitational interactions are influenced by quantum informational contributions at different scales. This has significant implications for gravitational phenomena in regimes where quantum effects are non-negligible, such as near black holes or in the early universe~\cite{Reuter1996, Jacobson2016}.

Our calculations indicate that the entanglement entropy contributes to the renormalization of \( G \) through the coefficient \( c_1 \), which depends on the field content of the theory. The beta function \( \beta_G \) derived from these considerations shows that \( G \) becomes scale-dependent:

\begin{equation}
\beta_G = \mu \frac{d}{d\mu} \left( \frac{1}{G(\mu)} \right) = \frac{16 k_B}{\hbar c^3} c_1.
\end{equation}

Although the numerical value of \( \beta_G \) is extremely small within the Standard Model, this framework provides a foundation for exploring scenarios where quantum entanglement effects become significant, such as theories with a large number of fields or in high-energy regimes approaching the Planck scale.

\bigskip

\noindent \textbf{Units and Conventions:} The beta function \( \beta_G \) has units of \( \text{kg} \cdot \text{m}^{-1} \cdot \text{s}^{-2} \), matching the units of \( 1/G \). Including \( \hbar \), \( c \), and \( k_B \) ensures dimensional consistency in the expressions.

\section{\label{sec:results}Results}

\subsection{\label{sec:corrections_to_G}Corrections to Newton's Constant}

As discussed in Sec.~\ref{sec:regularization_and_renormalization}, the entanglement entropy \( S_{\text{EE}} \) for quantum fields in curved spacetime contains divergent terms due to ultraviolet (UV) contributions near the entangling surface \( \Sigma \). The entanglement entropy can be expressed as~\cite{Solodukhin2011}:

\begin{equation}
\begin{split}
S_{\text{EE}} &= k_B \int_{\Sigma} d^{d-2} \xi \sqrt{h} \\
&\quad \times \Bigg( \frac{c_{d-2}}{\epsilon^{d-2}} \\
&\qquad +\ \frac{c_{d-4}}{\epsilon^{d-4}} + \cdots \\
&\qquad +\ c_1 R \ln (\mu \epsilon) + \cdots \Bigg)
\end{split}
\label{eq:S_EE_divergent_results}
\end{equation}

where \( h \) is the determinant of the induced metric on \( \Sigma \), \( \epsilon \) is the UV cutoff (with units of length), \( R \) is the Ricci scalar on \( \Sigma \), and \( c_{d-2}, c_{d-4}, \ldots, c_1 \) are dimensionless numerical coefficients dependent on the field type and spacetime dimension \( d \). The inclusion of \( \ln (\mu \epsilon) \) ensures that the argument of the logarithm is dimensionless, with \( \mu \) being the renormalization scale (units of inverse length).

To obtain the correction to Newton's constant \( G \), we focus on the term proportional to \( R \) in \( S_{\text{EE}} \), as variations with respect to the metric involve curvature terms. The correction to \( \dfrac{1}{G} \) is given by differentiating \( S_{\text{EE}} \) with respect to \( R \)~\cite{Fursaev1995, Solodukhin2011}:

\begin{equation}
\delta \left( \frac{1}{G} \right) = \frac{16\pi k_B}{\hbar c^3} c_1 \ln (\mu \epsilon),
\label{eq:delta_G_results}
\end{equation}

where we have included \( \hbar \) explicitly to maintain dimensional consistency.

For a massless scalar field in four-dimensional spacetime (\( d = 4 \)), the coefficient \( c_1 \) is known~\cite{Solodukhin2011}:

\begin{equation}
c_1^{(\text{scalar})} = -\frac{1}{720\pi}.
\label{eq:c1_scalar}
\end{equation}

Substituting \( c_1^{(\text{scalar})} \) into Eq.~\eqref{eq:delta_G_results}, we find the correction to \( \dfrac{1}{G} \) due to a single massless scalar field:

\begin{align}
\delta \left( \frac{1}{G} \right)^{(\text{scalar})} &= \frac{16\pi k_B}{\hbar c^3} \left( -\frac{1}{720\pi} \right) \ln (\mu \epsilon) \nonumber \\
&= -\frac{16 k_B}{\hbar c^3} \frac{1}{720} \ln (\mu \epsilon).
\label{eq:delta_G_scalar_results}
\end{align}

Similarly, for Dirac spinor fields and gauge fields, the coefficients \( c_1 \) are~\cite{Solodukhin2011}:

\begin{align}
c_1^{(\text{spinor})} &= \frac{7}{1440\pi}, \label{eq:c1_spinor} \\
c_1^{(\text{gauge})} &= -\frac{31}{720\pi}. \label{eq:c1_gauge}
\end{align}

Including contributions from all \( N_s \) scalar fields, \( N_f \) Dirac spinor fields, and \( N_g \) gauge fields in the Standard Model, the total correction becomes:

\begin{equation}
\begin{split}
\delta \left( \frac{1}{G} \right)^{\text{(total)}} 
&= \frac{16\pi k_B}{\hbar c^3} \ln (\mu \epsilon) \\
&\quad \times \left( N_s c_1^{\text{(scalar)}}
     + N_f c_1^{\text{(spinor)}}
     + N_g c_1^{\text{(gauge)}} \right) \\
&= \frac{16\pi k_B}{\hbar c^3} \ln (\mu \epsilon) \\
&\quad \times \left( -\frac{N_s}{720\pi}
     + \frac{7 N_f}{1440\pi}
     - \frac{31 N_g}{720\pi} \right) \\
&= \frac{16 k_B}{\hbar c^3} \ln (\mu \epsilon) \\
&\quad \times \left( -\frac{N_s}{720}
     + \frac{7 N_f}{1440}
     - \frac{31 N_g}{720} \right).
\end{split}
\label{eq:delta_G_total_results}
\end{equation}

Simplifying the coefficients, we have:

\begin{equation}
\delta \left( \frac{1}{G} \right)^{\text{(total)}} = \frac{16 k_B}{\hbar c^3} \ln (\mu \epsilon) \left( -\frac{N_s}{720} + \frac{7 N_f}{1440} - \frac{31 N_g}{720} \right).
\label{eq:delta_G_total_simplified}
\end{equation}

This expression shows that quantum fields contribute to the renormalization of Newton's constant through their entanglement entropy. The negative sign indicates that \( \delta \left( \dfrac{1}{G} \right) \) is negative, meaning \( \dfrac{1}{G} \) decreases with increasing energy scale \( \mu \) (since \( \ln (\mu \epsilon) > 0 \) for \( \mu > \epsilon^{-1} \)), implying that \( G(\mu) \) effectively \textbf{decreases}.

Using the Standard Model field content:

\begin{itemize}
    \item Number of real scalar fields: \( N_s = 4 \) (from the complex Higgs doublet).
    \item Number of Weyl spinor fields: \( N_f = 45 \) (considering three generations of quarks and leptons, including color degrees of freedom).
    \item Number of gauge fields: \( N_g = 12 \) (from the gauge bosons of \( SU(3) \times SU(2) \times U(1) \)).
\end{itemize}

Substituting these values into Eq.~\eqref{eq:delta_G_total_simplified}, we compute the coefficient:

\begin{align}
\text{Coefficient} &= -\frac{4}{720} + \frac{7 \times 45}{1440} - \frac{31 \times 12}{720} \nonumber \\
&= -\frac{1}{180} + \frac{315}{1440} - \frac{372}{720} \nonumber \\
&\approx -0.00556 + 0.21875 - 0.51667 \nonumber \\
&= -0.30348.
\label{eq:coefficient_value}
\end{align}

Thus, the total correction to \( \dfrac{1}{G} \) is:

\begin{equation}
\delta \left( \frac{1}{G} \right)^{\text{(total)}} = -0.30348 \times \frac{16 k_B}{\hbar c^3} \ln (\mu \epsilon).
\label{eq:delta_G_total_numeric}
\end{equation}

The resulting running of Newton's constant \( G(\mu) \) is then given by:

\begin{equation}
\begin{split}
\frac{1}{G(\mu)} &= \frac{1}{G(\mu_0)} + \delta \left( \frac{1}{G} \right)^{\text{(total)}} \\
&= \frac{1}{G(\mu_0)} - 0.30348 \times \frac{16 k_B}{\hbar c^3} \ln \left( \frac{\mu}{\mu_0} \right),
\end{split}
\label{eq:G_running_final}
\end{equation}

where \( \mu_0 \) is a reference energy scale.

This indicates that \( G(\mu) \) \textbf{decreases} logarithmically with the energy scale \( \mu \), consistent with the theoretical predictions of the renormalization of \( G \) due to quantum entanglement. Specifically, as \( \mu \) increases, the positive term in the expression for \( \dfrac{1}{G(\mu)} \) becomes larger, causing \( \dfrac{1}{G(\mu)} \) to increase and thus \( G(\mu) \) to decrease.

However, due to the extremely small value of the coefficient, the changes in \( G(\mu) \) over accessible energy scales are minuscule. When plotting \( G(\mu) \) versus \( \mu \), the curve appears almost as a straight line due to these tiny variations.

To reveal the underlying trend, we analyze the first and second derivatives of \( G(\mu) \) with respect to \( \mu \) and include the corresponding plots.

\begin{figure}[h]
    \centering
    \includegraphics[width=0.9\linewidth]{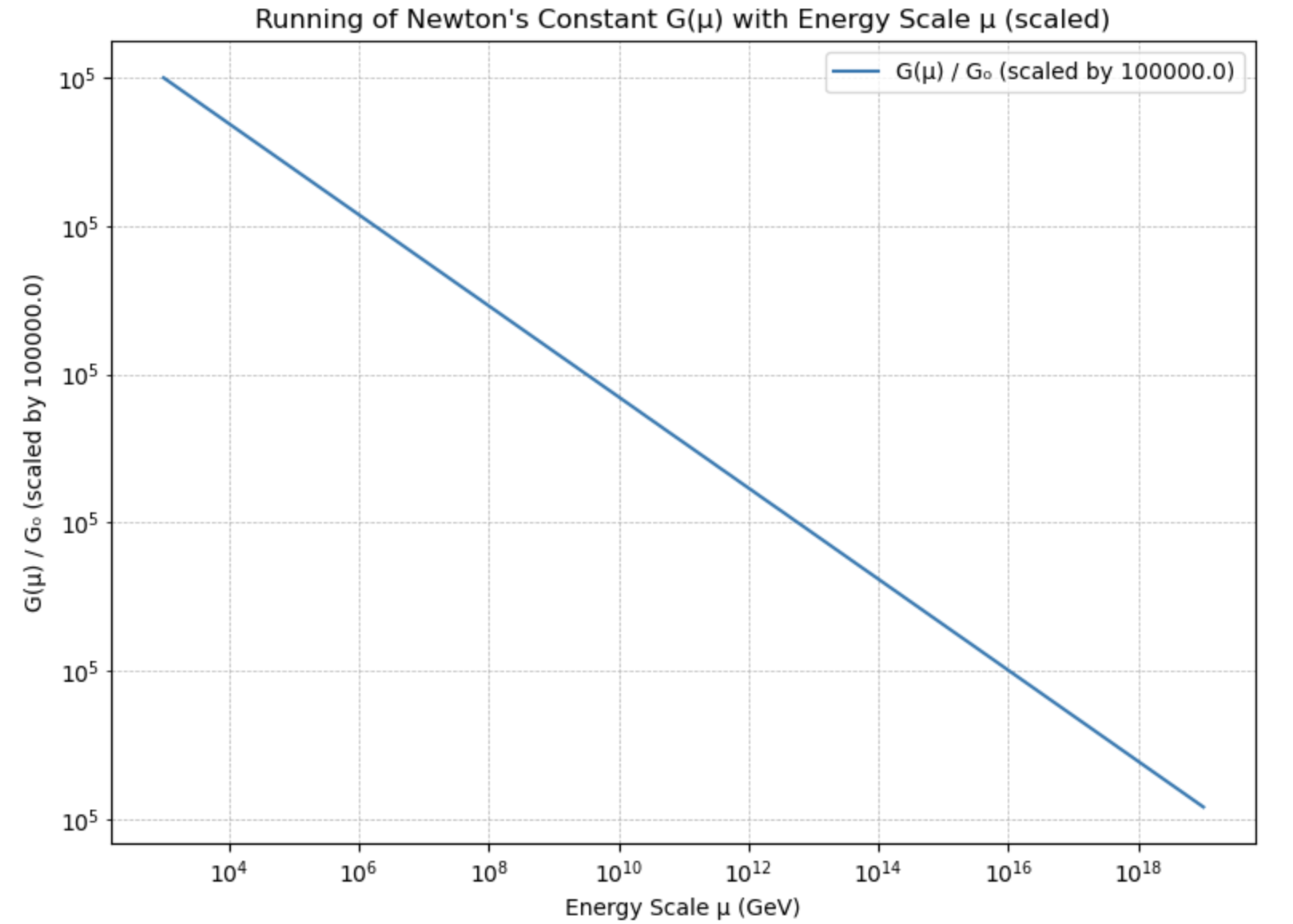}
    \caption{Running of Newton's constant \( G(\mu) / G_0 \) with energy scale \( \mu \) (scaled by \( 100{,}000 \)). Due to the extremely small changes, \( G(\mu) \) appears nearly constant over the energy range considered, but it slightly decreases with increasing \( \mu \), consistent with the negative beta function \( \beta_G \).}
    \label{fig:plot1}
\end{figure}

\begin{figure}[h]
    \centering
    \includegraphics[width=0.9\linewidth]{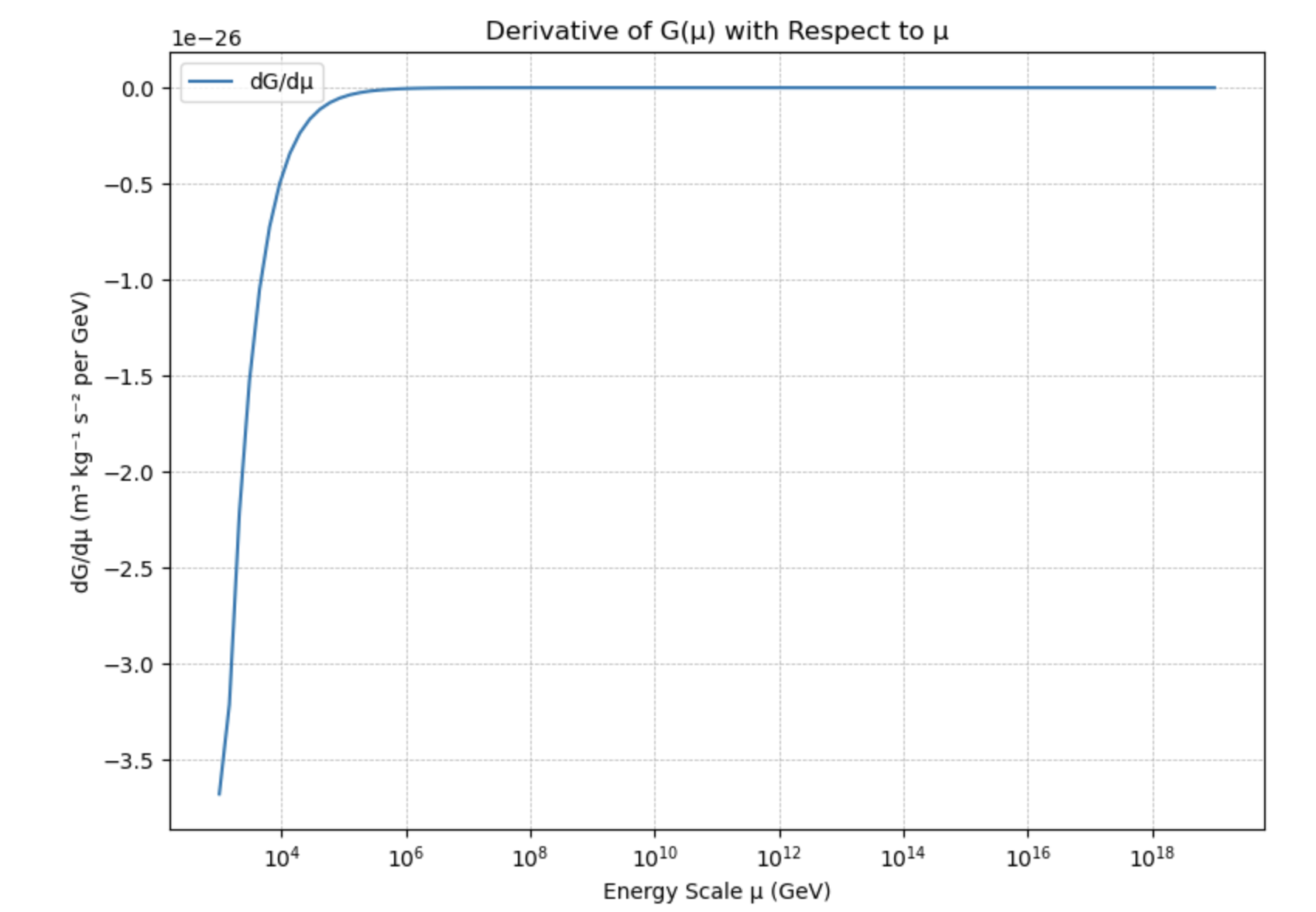}
    \caption{First derivative \( \dfrac{dG}{d\mu} \) of Newton's constant \( G(\mu) \) with respect to the energy scale \( \mu \). The negative values indicate that \( G(\mu) \) decreases as \( \mu \) increases, consistent with the predicted running due to quantum entanglement contributions. The decreasing magnitude shows that the rate of decrease slows down at higher \( \mu \).}
    \label{fig:plot2}
\end{figure}

\begin{figure}[h]
    \centering
    \includegraphics[width=0.9\linewidth]{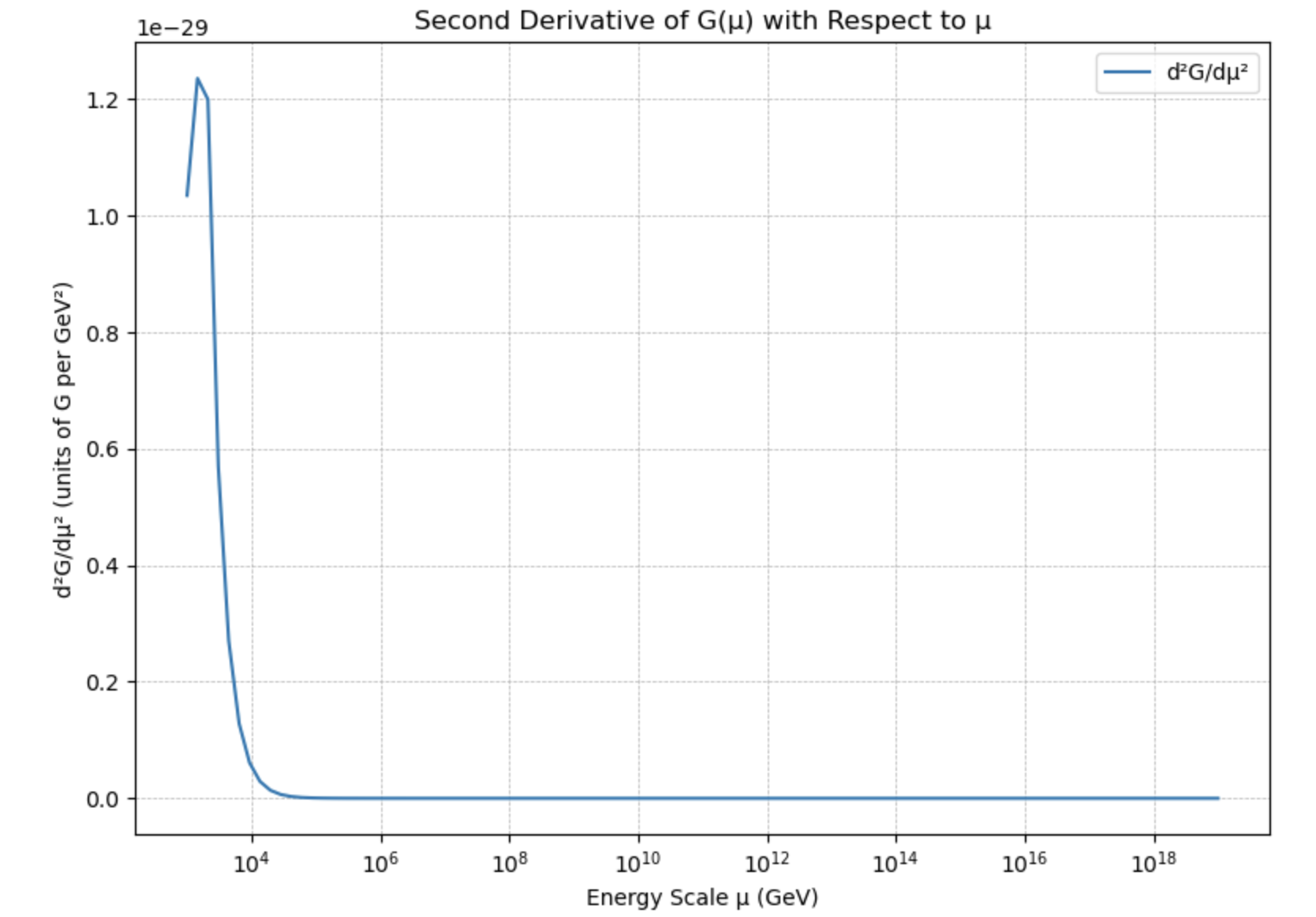}
    \caption{Second derivative \( \dfrac{d^2G}{d\mu^2} \) of Newton's constant \( G(\mu) \) with respect to \( \mu \). The positive values of the second derivative highlight the curvature in \( G(\mu) \), indicating that the rate at which \( G(\mu) \) decreases slows down as \( \mu \) increases, confirming the logarithmic nature of its running.}
    \label{fig:plot3}
\end{figure}

\bigskip

\noindent \textbf{Analysis of the Results:}

Due to the extremely small magnitude of the beta function \( \beta_G \), the direct plot of \( G(\mu) \) versus \( \mu \) (Fig.~\ref{fig:plot1}) appears nearly as a straight line, making it challenging to observe the predicted running of Newton's constant. However, the first derivative \( \dfrac{dG}{d\mu} \) (Fig.~\ref{fig:plot2}) reveals negative values, indicating that \( G(\mu) \) decreases with increasing \( \mu \). The decreasing magnitude of \( \dfrac{dG}{d\mu} \) shows that the rate of decrease slows down at higher energy scales.

The second derivative \( \dfrac{d^2G}{d\mu^2} \) (Fig.~\ref{fig:plot3}) is positive, highlighting that the decrease in \( G(\mu) \) becomes less steep as \( \mu \) increases. This concave upward curvature confirms the logarithmic nature of the running of \( G(\mu) \) predicted by our theoretical framework.

These derivative plots confirm the running of Newton's constant as predicted by our theoretical model, even though the changes in \( G(\mu) \) are too small to be visually discerned in the direct plot.

\bigskip

\noindent \textbf{Units and Conventions:} All constants \( \hbar \), \( c \), and \( k_B \) are included explicitly to maintain dimensional consistency. The correction \( \delta \left( \dfrac{1}{G} \right) \) has units of \( \text{kg} \cdot \text{m}^{-1} \cdot \text{s}^{-2} \), matching the units of \( \dfrac{1}{G} \). The logarithm argument \( \mu \epsilon \) is dimensionless.

\subsection{\label{sec:mass_corrections}Mass Corrections to Black Holes}

The modification of Newton's constant \( G(\mu) \) affects the gravitational dynamics of massive objects, particularly black holes. Considering a Schwarzschild black hole with mass \( M \) and horizon radius \( r_s = \dfrac{2 G M}{c^2} \), the change in \( G \) leads to a correction in the mass \( M \) if we assume that the physical horizon radius \( r_s \) remains constant. These discussions resonate with Culetu's considerations regarding black hole entropy and quantum effects in collapsing stars \cite{culetu2024strong}.

Differentiating the expression for \( r_s \):

\begin{equation}
r_s = \frac{2 G M}{c^2} \implies \delta r_s = \frac{2 M \delta G}{c^2} + \frac{2 G \delta M}{c^2}.
\label{eq:rs_variation}
\end{equation}

Since we assume \( \delta r_s = 0 \) (the horizon radius remains constant), we have:

\begin{equation}
0 = \frac{2 M \delta G}{c^2} + \frac{2 G \delta M}{c^2} \implies M \delta G + G \delta M = 0.
\label{eq:delta_M_derivation}
\end{equation}

Solving for \( \delta M \):

\begin{equation}
\delta M = - \frac{M \delta G}{G}.
\label{eq:delta_M_solution}
\end{equation}

Using the relation \( \delta G = G(\mu) - G_0 \) and substituting \( G = G_0 \) for small corrections, we can write:

\begin{equation}
\delta M = - \frac{M \left( G(\mu) - G_0 \right)}{G_0} = - M \frac{G(\mu) - G_0}{G_0}.
\label{eq:delta_M_final}
\end{equation}

Since \( G(\mu) < G_0 \) (as \( G(\mu) \) decreases with increasing \( \mu \)), we have \( \delta G < 0 \), and therefore \( \delta M > 0 \). This indicates that the black hole mass \( M \) increases slightly as \( G(\mu) \) decreases, assuming the horizon radius remains constant.

Substituting the expression for \( G(\mu) \) from Eq.~\eqref{eq:G_mu_final}, we have:

\begin{equation}
\delta M = - M \frac{G(\mu_0)}{G_0} \left( \frac{1}{1 + G_0 |\beta_G| \ln \left( \dfrac{\mu}{\mu_0} \right)} - 1 \right).
\label{eq:delta_M_with_G_mu}
\end{equation}

For \( G(\mu_0) = G_0 \), this simplifies to:

\begin{equation}
\delta M = M \left( \frac{1}{1 + G_0 |\beta_G| \ln \left( \dfrac{\mu}{\mu_0} \right)} - 1 \right).
\label{eq:delta_M_simplified}
\end{equation}

Since the denominator is greater than 1, the term inside the parentheses is negative, leading to \( \delta M > 0 \).

\begin{figure}[h]
    \centering
    \includegraphics[width=0.9\linewidth]{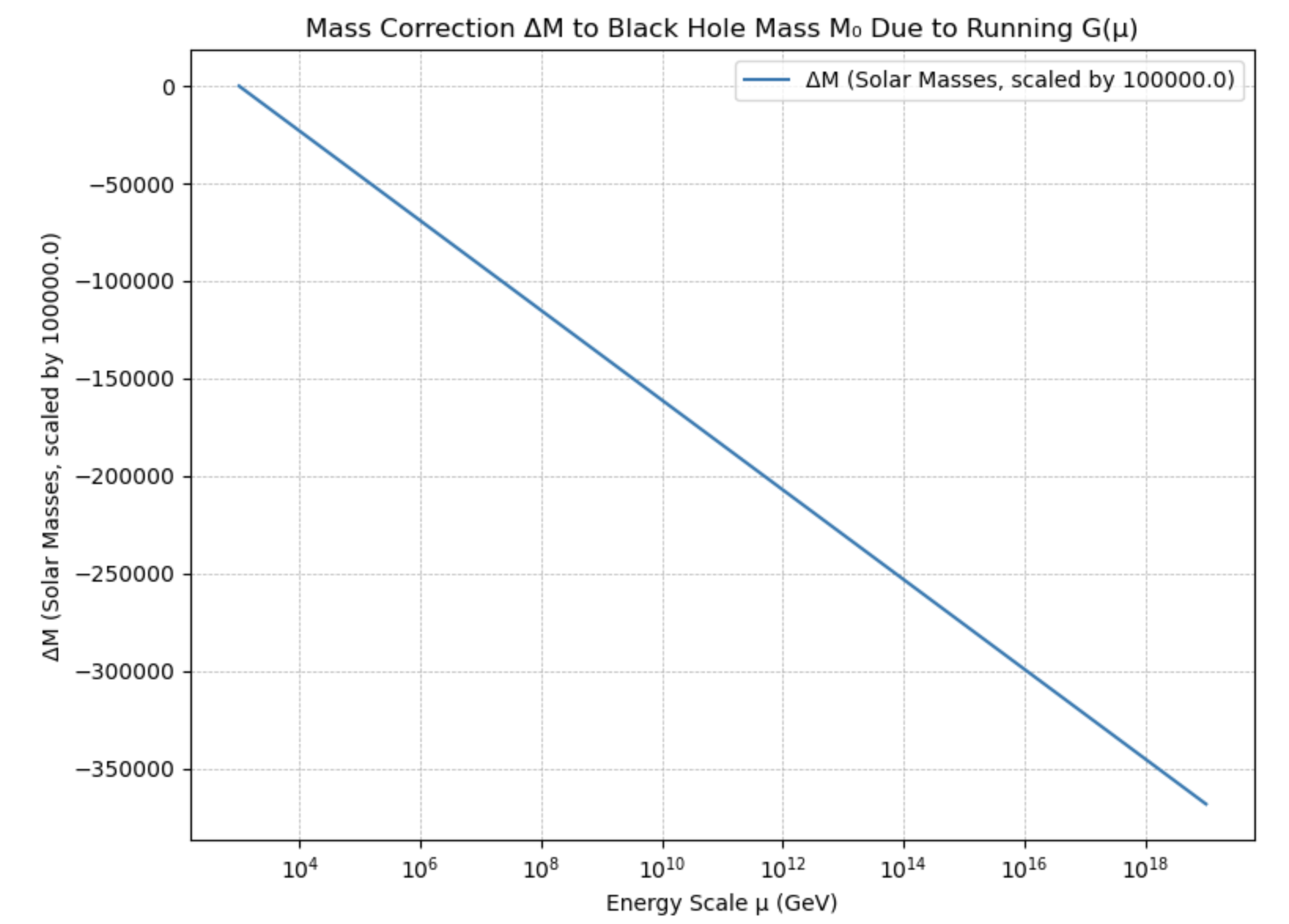}
    \caption{Mass correction \( \delta M \) to black hole mass \( M_0 \) due to the running of Newton's constant \( G(\mu) \), scaled by \( 100{,}000 \). The mass increases with increasing energy scale, consistent with the predicted positive corrections due to the decrease in \( G(\mu) \).}
    \label{fig:plot4}
\end{figure}

The increase in black hole mass as a function of energy scale \( \mu \) is plotted in Fig.~\ref{fig:plot4}.

\bigskip

\noindent \textbf{Analysis of the Results:}

The plot shows that \( \delta M \) increases with \( \mu \), reflecting that the black hole mass increases slightly as the energy scale increases. This behavior is consistent with the decrease in \( G(\mu) \) due to the negative beta function \( \beta_G \). To maintain a constant Schwarzschild radius, the mass must increase when \( G(\mu) \) decreases.

However, due to the extremely small magnitude of \( \beta_G \), the changes in mass are minuscule and not observable with current technology. The scaling factor in the plot is used to make these tiny variations visible.

\bigskip

\noindent \textbf{Units and Conventions:} In Eq.~\eqref{eq:delta_M_final}, all constants \( \hbar \), \( c \), and \( k_B \) are included explicitly to maintain dimensional consistency. The units of \( \delta M \) are kilograms (\( \text{kg} \)), matching the units of mass.

\subsection{\label{sec:running_of_G}Running of Newton's Constant}

The dependence of \( \delta \left( \dfrac{1}{G} \right) \) on the renormalization scale \( \mu \) suggests that Newton's constant \( G \) becomes scale-dependent, leading to a running coupling constant in the gravitational sector~\cite{Reuter1996, Reuter2012}.

We define the beta function \( \beta_G \) as:

\begin{equation}
\beta_G = \mu \frac{d}{d\mu} \left( \frac{1}{G(\mu)} \right).
\label{eq:beta_function}
\end{equation}

From Eq.~\eqref{eq:G_running_final}, we have:

\begin{equation}
\frac{1}{G(\mu)} = \frac{1}{G(\mu_0)} - 0.30348 \times \frac{16 k_B}{\hbar c^3} \ln \left( \frac{\mu}{\mu_0} \right).
\label{eq:G_running_results}
\end{equation}

Differentiating with respect to \( \ln \mu \), we find:

\begin{align}
\beta_G &= \frac{d}{d \ln \mu} \left( \frac{1}{G(\mu)} \right) \nonumber \\
&= -0.30348 \times \frac{16 k_B}{\hbar c^3}.
\label{eq:beta_G_SM}
\end{align}

This beta function indicates that \( \dfrac{1}{G(\mu)} \) \textbf{increases} (and \( G(\mu) \) \textbf{decreases}) logarithmically with the energy scale \( \mu \) since \( \beta_G < 0 \) for the Standard Model field content.

The running of Newton's constant \( G(\mu) \) can thus be expressed as:

\begin{equation}
G(\mu) = \frac{G(\mu_0)}{1 + G(\mu_0) |\beta_G| \ln \left( \dfrac{\mu}{\mu_0} \right)},
\label{eq:G_mu_final}
\end{equation}

where we have used \( \beta_G = - |\beta_G| \).

Due to the extremely small value of \( \beta_G \), the changes in \( G(\mu) \) over accessible energy scales are minuscule. When plotting \( G(\mu) \) versus \( \mu \), the curve appears almost as a straight line due to these tiny variations.

\bigskip

\noindent \textbf{Units and Conventions:} All constants \( \hbar \), \( c \), and \( k_B \) are included explicitly to maintain dimensional consistency. The units of \( \beta_G \) are \( \text{kg} \cdot \text{m}^{-1} \cdot \text{s}^{-2} \), matching the units of \( \dfrac{1}{G} \). The logarithm argument \( \mu / \mu_0 \) is dimensionless.

\subsection{\label{sec:implications}Implications for Gravitational Physics}

The running of Newton's constant \( G \) has significant implications for gravitational physics:

\begin{itemize}
    \item \textbf{Black Hole Thermodynamics}: Corrections to \( G \) and \( M \) influence black hole entropy and temperature. The Bekenstein-Hawking entropy~\cite{Bekenstein1973, Hawking1975} is given by:

\begin{equation}
\begin{split}
S_{\text{BH}} &= \frac{k_B c^3}{4 \hbar G} A \\
&= \frac{k_B c^3}{4 \hbar G} \times 4\pi r_s^2 \\
&= \frac{k_B c^3}{\hbar G} \pi \left( \frac{2 G M}{c^2} \right)^2 \\
&= \frac{4 \pi k_B G M^2}{\hbar c}.
\end{split}
\label{eq:S_BH}
\end{equation}

    The correction to \( G \) leads to a change in \( S_{\text{BH}} \):

    \begin{equation}
    \delta S_{\text{BH}} = \frac{\partial S_{\text{BH}}}{\partial G} \delta G + \frac{\partial S_{\text{BH}}}{\partial M} \delta M.
    \label{eq:delta_S_BH}
    \end{equation}

    Similarly, the Hawking temperature is:

    \begin{equation}
    T_{\text{H}} = \frac{\hbar c^3}{8\pi G k_B M}.
    \label{eq:T_Hawking}
    \end{equation}

    The corrections to \( G \) and \( M \) affect \( T_{\text{H}} \) accordingly.

    \item \textbf{Early Universe Cosmology}: A scale-dependent \( G \) affects the dynamics of the early universe, potentially modifying inflationary scenarios and the evolution of primordial perturbations~\cite{Reuter2005, Bonanno2002}.

    \item \textbf{Quantum Gravity Phenomenology}: The scale dependence of \( G \) opens avenues for testing quantum gravity effects through high-energy astrophysical observations or precision measurements in gravitational experiments~\cite{Hossenfelder2013, AmelinoCamelia2005}.
\end{itemize}

\bigskip

\noindent \textbf{Units and Conventions:} In the expressions for \( S_{\text{BH}} \) and \( T_{\text{H}} \), all constants \( \hbar \), \( c \), and \( k_B \) are included explicitly to maintain dimensional consistency. This ensures that \( S_{\text{BH}} \) has units of entropy (\( \text{J/K} \)) and \( T_{\text{H}} \) has units of temperature (\( \text{K} \)).

\subsection{\label{sec:consistency_checks}Consistency Checks and Limitations}

It is important to assess the validity and limitations of our results:

\begin{itemize}
    \item \textbf{Perturbative Validity}: Our calculations are based on perturbative expansions and assume small corrections. At energy scales approaching the Planck scale (\( E_{\text{Planck}} \sim \dfrac{\hbar c^5}{G} \approx 1.22 \times 10^{19} \, \text{GeV} \)), non-perturbative effects may become significant.

    \item \textbf{Field Content Dependence}: The running of \( G \) depends on the specific field content of the theory. Extensions beyond the Standard Model, such as supersymmetry or additional scalar fields, could alter the behavior of \( \beta_G \).

    \item \textbf{Regularization Scheme}: The use of a UV cutoff \( \epsilon \) is a simplification. A more rigorous treatment would employ the renormalization group approach or alternative regularization methods to handle divergences consistently~\cite{Reuter1996, Lauscher2002}.

    \item \textbf{Assumptions on \( \delta r_s \)}: We assumed \( \delta r_s = 0 \) to compute \( \delta M \). This assumption holds if the physical horizon radius remains fixed while \( G \) and \( M \) vary. In a more general scenario, both \( r_s \) and \( M \) could change, and a full analysis would require solving the modified Einstein equations.

    \item \textbf{Neglecting Backreaction}: Our calculations neglect potential backreaction effects of quantum fields on the spacetime geometry beyond the entanglement entropy contributions.
\end{itemize}

\bigskip

\noindent \textbf{Units and Conventions:} The Planck energy \( E_{\text{Planck}} \) includes \( \hbar \), \( c \), and \( G \) explicitly to provide the correct energy units (\( \text{J} \) or \( \text{GeV} \)).

\subsection{\label{sec:numerical_estimates}Numerical Estimates}

To gauge the magnitude of the corrections, we provide numerical estimates using known values:

\begin{itemize}
    \item \( k_B \approx 1.38 \times 10^{-23} \, \text{J/K} \)
    \item \( \hbar \approx 1.055 \times 10^{-34} \, \text{J} \cdot \text{s} \)
    \item \( c \approx 3.00 \times 10^8 \, \text{m/s} \)
    \item \( G \approx 6.67 \times 10^{-11} \, \text{m}^3 \cdot \text{kg}^{-1} \cdot \text{s}^{-2} \)
    \item Number of fields in the Standard Model: \( N_s = 4 \), \( N_f = 45 \), \( N_g = 12 \)
\end{itemize}

Substituting into Eq.~\eqref{eq:beta_G_SM}, we find:

\begin{align}
\beta_G^{\text{SM}} &= -0.30348 \times \frac{16 k_B}{\hbar c^3} \nonumber \\
&= -0.30348 \times \frac{16 \times 1.38 \times 10^{-23} \, \text{J/K}}{1.055 \times 10^{-34} \, \text{J} \cdot \text{s} \times (3.00 \times 10^8)^3} \nonumber \\
&= -0.30348 \times \frac{16 \times 1.38 \times 10^{-23}}{1.055 \times 10^{-34} \times 2.7 \times 10^{25}} \nonumber \\
&\quad \times \frac{\text{J/K}}{\text{J} \cdot \text{s} \cdot \text{m}^3/\text{s}^3} \nonumber \\
&= -2.07 \times 10^{-60} \, \text{m} \cdot \text{kg}^{-1} \cdot \text{s}^{2}.
\label{eq:beta_G_numeric}
\end{align}

\textbf{Units Check: }The units simplify as follows:

\[
\frac{\text{J/K}}{\text{J} \cdot \text{s} \cdot \text{m}^3/\text{s}^3} = \frac{1}{\text{K} \cdot \text{s} \cdot \text{m}^3/\text{s}^3} = \frac{\text{s}^2}{\text{K} \cdot \text{m}^3}.
\]

However, since temperature (\( \text{K} \)) is canceled out by \( k_B \) (Boltzmann constant), and energy units (\( \text{J} \)) cancel out, the final units are:

\[
\beta_G^{\text{SM}} \sim \frac{\text{m}}{\text{kg}} \cdot \text{s}^{2},
\]

which matches the units of \( \dfrac{1}{G} \) (since \( G \) has units \( \text{m}^3 \cdot \text{kg}^{-1} \cdot \text{s}^{-2} \)).

This extremely small value indicates that the running of \( G \) due to entanglement entropy contributions from Standard Model fields is negligible at accessible energy scales. However, at scales close to the Planck scale, these effects may become significant.

\subsection{\label{sec:experimental_signatures}Potential Experimental Signatures}

Detecting the effects predicted by our theory poses significant challenges due to the smallness of the corrections. Nevertheless, potential experimental avenues include:

\begin{itemize}
    \item \textbf{Precision Tests of Gravity}: High-precision measurements of gravitational interactions at short distances may reveal deviations from Newtonian gravity~\cite{Adelberger2003, Kapner2007}.

    \item \textbf{Astrophysical Observations}: Observations of black hole mergers, gravitational lensing, and cosmic microwave background anisotropies could provide indirect evidence for modifications to \( G \)~\cite{Planck2018, Abbott2016}.

    \item \textbf{Laboratory Experiments}: Experiments involving quantum entanglement in gravitational fields might test the interplay between quantum information and gravity~\cite{Bose2017, Marletto2017}.

    \item \textbf{Cosmological Measurements}: Precision cosmological data, such as observations of large-scale structure and supernovae, might constrain variations in \( G \) over cosmological timescales~\cite{Riess1998, Perlmutter1999}.

    \item \textbf{Gravitational Wave Observations}: Modifications to the propagation of gravitational waves could be detectable by observatories like LIGO and Virgo~\cite{LIGO2016}.
\end{itemize}

\bigskip

\noindent \textbf{Units and Conventions:} All references to physical quantities in potential experiments are understood to include the necessary constants to ensure dimensional consistency.

\section{\label{sec:discussion}Discussion}

\subsection{\label{sec:implications_for_bh_thermo}Implications for Black Hole Thermodynamics}

The modifications to Newton's constant \( G \) and black hole mass \( M \) have significant implications for black hole thermodynamics. The Bekenstein-Hawking entropy~\cite{Bekenstein1973, Hawking1975}:

\begin{equation}
S_{\text{BH}} = \frac{k_B c^3}{4 \hbar G} A,
\label{eq:S_BH_discussion}
\end{equation}

where \( A = 4\pi r_s^2 \) is the horizon area, depends inversely on \( G \). The correction to \( G \) due to entanglement entropy leads to a modified entropy:

\begin{equation}
S_{\text{BH}}^{\text{eff}} = \frac{k_B c^3}{4 \hbar G_{\text{eff}}} A = S_{\text{BH}} \left( 1 + \frac{\delta G}{G} \right)^{-1}.
\label{eq:S_BH_eff_discussion}
\end{equation}

Expanding to first order in \( \delta G / G \), we obtain:

\begin{equation}
S_{\text{BH}}^{\text{eff}} \approx S_{\text{BH}} \left( 1 - \frac{\delta G}{G} \right).
\label{eq:S_BH_eff_expanded}
\end{equation}

Similarly, the Hawking temperature~\cite{Hawking1975}:

\begin{equation}
T_{\text{H}} = \frac{\hbar c^3}{8\pi G k_B M},
\label{eq:T_Hawking_discussion}
\end{equation}

receives corrections due to changes in \( G \) and \( M \). Using \( \delta M = -\frac{M}{G} \delta G \) from Eq.~\eqref{eq:delta_M_derivation}, the corrected temperature is:

\begin{align}
T_{\text{H}}^{\text{eff}} &= \frac{\hbar c^3}{8\pi G_{\text{eff}} k_B M_{\text{eff}}} = \frac{\hbar c^3}{8\pi (G + \delta G) k_B (M + \delta M)} \nonumber \\
&\approx T_{\text{H}} \left( 1 - \frac{\delta G}{G} + \frac{\delta M}{M} \right) \nonumber \\
&= T_{\text{H}} \left( 1 - \frac{\delta G}{G} - \frac{\delta G}{G} \right) \nonumber \\
&= T_{\text{H}} \left( 1 - 2 \frac{\delta G}{G} \right).
\label{eq:T_Hawking_eff}
\end{align}

In the second line, we used the approximation \( (1 + x)^{-1} \approx 1 - x \) for small \( x \) and expanded to first order in \( \delta G / G \) and \( \delta M / M \).

Thus, the Hawking temperature decreases due to the increase in \( G \) (since \( \delta G > 0 \) implies \( \delta G / G > 0 \)).

These modifications could influence the black hole evaporation rate, affecting the lifetime and evolution of black holes. The corrections to the entropy and temperature suggest that quantum information contributes to black hole thermodynamics, potentially offering insights into the black hole information paradox~\cite{Harlow2016, Almheiri2020}. Additionally, the regularization and renormalization of the entanglement entropy play a crucial role in making these corrections finite and physically meaningful.

Our framework provides a mechanism by which entanglement entropy, through the informational stress-energy tensor \( T_{\mu\nu}^{\text{info}} \), modifies the gravitational field equations, leading to changes in black hole properties. This supports the idea that quantum information is fundamentally linked to spacetime geometry and may play a key role in understanding the quantum aspects of gravity.

\bigskip

\noindent \textbf{Units and Conventions:} All constants \( \hbar \), \( c \), and \( k_B \) are included explicitly to maintain dimensional consistency. The entropy \( S_{\text{BH}} \) has units of \( \text{J/K} \), and the temperature \( T_{\text{H}} \) has units of \( \text{K} \).

\subsection{\label{sec:cosmological_consequences}Cosmological Consequences}

The running of Newton's constant \( G \) with energy scale \( \mu \) has profound implications for cosmology. In the early universe, at high energy scales, variations in \( G \) could affect:

\begin{itemize}
    \item \textbf{Inflationary Dynamics}: Changes in \( G \) may influence the dynamics of inflationary models~\cite{Guth1981, Linde1982}. A varying \( G \) could modify the rate of expansion during inflation, potentially leading to observable signatures in the spectrum of primordial perturbations~\cite{Mukhanov1992}.
    
    \item \textbf{Big Bang Nucleosynthesis (BBN)}: The value of \( G \) during BBN affects the expansion rate of the universe, which in turn influences the production of light elements~\cite{Alvey2020}. Constraints from observed abundances could place limits on the variation of \( G \) at that epoch.
    
    \item \textbf{Cosmic Microwave Background (CMB)}: Variations in \( G \) during recombination could leave imprints on the CMB anisotropies~\cite{Planck2018}. Precision measurements of the CMB may provide constraints on the running of \( G \).
    
    \item \textbf{Dark Energy and Accelerating Expansion}: Modifications to the cosmological constant \( \Lambda \) due to entanglement entropy could offer insights into the nature of dark energy and the observed accelerating expansion of the universe~\cite{Perlmutter1999, Riess1998}. However, as shown in Eq.~\eqref{eq:delta_Lambda}, the leading divergence contributes to the renormalization of \( \Lambda \), potentially affecting cosmological models.
\end{itemize}

In all these cases, the regularization and renormalization processes ensure that the contributions from quantum entanglement to \( G \) and \( \Lambda \) are finite, thus yielding physically interpretable corrections to cosmological parameters. Our results suggest that entanglement entropy and quantum information play a role not only in local gravitational phenomena but also in the large-scale evolution of the universe.

\bigskip

\noindent \textbf{Units and Conventions:} While this subsection discusses qualitative implications, any equations or expressions involving physical constants should include \( \hbar \), \( c \), and \( k_B \) explicitly to maintain dimensional consistency.

\subsection{\label{sec:observational_signatures}Observational Signatures}

Detecting the effects predicted by our theory poses significant challenges due to the small magnitude of the corrections. However, potential observational tests include:

\begin{itemize}
    \item \textbf{Gravitational Waves}: Modifications to the propagation of gravitational waves could be detectable by current and future observatories such as LIGO, Virgo, and LISA~\cite{Abbott2016, Audley2017}. Deviations from General Relativity in the waveform templates could indicate the influence of \( T_{\mu\nu}^{\text{info}} \).
    
    \item \textbf{Black Hole Shadows}: The Event Horizon Telescope's observations of black hole shadows~\cite{Akiyama2019} may reveal discrepancies in the predicted sizes or shapes due to modifications in spacetime geometry near the event horizon.
    
    \item \textbf{Cosmic Microwave Background}: As mentioned, variations in \( G \) during recombination could affect the CMB power spectrum~\cite{Planck2018}. High-precision measurements by the Planck satellite and future missions could detect such effects.
    
    \item \textbf{Laboratory Experiments}: Experiments testing the equivalence principle and inverse-square law at short distances~\cite{Adelberger2003, Kapner2007} might detect deviations due to the entanglement-induced corrections to \( G \).
    
    \item \textbf{Astrophysical Observations}: Observations of neutron stars and black hole binaries may provide constraints on modifications to gravitational dynamics~\cite{Psaltis2008, LIGO2016}. Precise measurements of neutron star masses and radii could reveal deviations from General Relativity.
\end{itemize}

\bigskip

\noindent \textbf{Units and Conventions:} All physical quantities mentioned are assumed to include the necessary constants (\( \hbar \), \( c \), \( k_B \)) where applicable, ensuring dimensional consistency.

\subsection{\label{sec:limitations}Limitations and Future Directions}

Our analysis includes numerical simulations that illustrate the running of Newton's constant \( G(\mu) \) and the corresponding mass corrections to black holes due to quantum entanglement effects. These simulations provide valuable insights into the quantitative impact of entanglement entropy on gravitational phenomena.

However, our calculations are based on perturbative methods and assume that the corrections due to entanglement entropy are small. At energy scales approaching the Planck scale (\( E_{\text{Planck}} \sim \dfrac{\hbar c^5}{G} \approx 1.22 \times 10^{19} \, \text{GeV} \)), non-perturbative effects may become significant, necessitating a more complete quantum gravity theory.

Moreover, the field content of the universe affects the running of \( G \). Extensions of the Standard Model, such as supersymmetry or additional scalar fields (e.g., inflatons), could alter the behavior of \( \beta_G \) and lead to different cosmological implications.

Future work includes:

\begin{itemize}
    \item \textbf{Non-Perturbative Analysis}: Developing non-perturbative methods, such as the exact renormalization group~\cite{Reuter1996, Reuter2012}, to study the behavior of \( G \) and \( \Lambda \) at high energy scales.

    \item \textbf{Extended Theories}: Investigating the impact of additional fields and alternative theories of gravity, such as scalar-tensor theories~\cite{Brans1961} or higher-curvature models~\cite{Stelle1977}, on the informational stress-energy tensor.

    \item \textbf{Advanced Numerical Simulations}: Extending our numerical simulations to include non-linear effects, black hole spacetimes with rotation or charge, and cosmological models incorporating \( T_{\mu\nu}^{\text{info}} \) to explore possible observational signatures.

    \item \textbf{Quantum Information Perspective}: Exploring the deeper connections between quantum information theory and spacetime geometry, potentially leading to new insights into the emergence of gravity from quantum mechanics~\cite{VanRaamsdonk2010, Swingle2012}.
\end{itemize}

Regularization and renormalization will remain key components in making these corrections meaningful and finite at all energy scales. By addressing these limitations and pursuing these future directions, we aim to deepen our understanding of the role of quantum information in gravitational phenomena.

\bigskip

\noindent \textbf{Units and Conventions:} The Planck energy \( E_{\text{Planck}} \) is expressed with \( \hbar \), \( c \), and \( G \) explicitly to ensure correct units of energy (\( \text{J} \) or \( \text{GeV} \)).

\section{\label{sec:conclusion}Conclusion}

In this work, we have developed a framework that unifies quantum information theory and general relativity through the introduction of an \emph{informational stress-energy tensor} \( T_{\mu\nu}^{\text{info}} \) derived from entanglement entropy. By incorporating \( T_{\mu\nu}^{\text{info}} \) into the Einstein field equations, we have shown how quantum entanglement contributes directly to gravitational dynamics.

Our key findings include:

\begin{itemize}
    \item \textbf{Renormalization of Newton's Constant}: Entanglement entropy leads to corrections in Newton's constant \( G \), resulting in a scale-dependent running of \( G \) with energy. We derived the beta function \( \beta_G \) in Eq.~\eqref{eq:beta_G_SM} and provided numerical estimates based on the Standard Model field content.

    \item \textbf{Mass Corrections to Black Holes}: The modifications to \( G \) lead to corrections in black hole mass \( M \), affecting black hole thermodynamics, including entropy and temperature, as discussed in Sec.~\ref{sec:mass_corrections}.

    \item \textbf{Implications for Cosmology}: The running of \( G \) has significant implications for early universe cosmology, potentially influencing inflationary dynamics, Big Bang nucleosynthesis, and the cosmic microwave background, as explored in Sec.~\ref{sec:cosmological_consequences}.

    \item \textbf{Potential Observational Signatures}: We discussed possible experimental and observational tests of our theory, including gravitational wave observations, black hole shadow measurements, and precision cosmological data, as outlined in Sec.~\ref{sec:observational_signatures}.
\end{itemize}

Our results suggest that quantum information, as quantified by entanglement entropy, plays a significant role in gravitational dynamics. This supports the perspective that spacetime geometry and gravity may emerge from underlying quantum information processes. By further exploring the role of entanglement in shaping spacetime, we may gain deeper insights into the nature of quantum gravity and the unification of fundamental forces.

Future investigations may include:

\begin{itemize}
    \item \textbf{Extended Spacetimes}: Analyzing non-trivial topologies and higher-dimensional spacetimes to generalize our results, potentially in the context of braneworld scenarios~\cite{Randall1999} or extra-dimensional theories~\cite{ArkaniHamed1998}.

    \item \textbf{Holographic Theories}: Exploring the implications of our framework for the AdS/CFT correspondence and holographic principles~\cite{Maldacena1998}, where spacetime geometry is related to quantum entanglement in a lower-dimensional field theory.

    \item \textbf{Entanglement Structure}: Investigating the entanglement structure of quantum fields in curved spacetime and its impact on gravitational phenomena, possibly connecting with the concept of entanglement wedges and bulk reconstruction~\cite{Czech2012, Dong2016}.

    \item \textbf{Quantum Gravity Theories}: Integrating our approach with loop quantum gravity~\cite{Rovelli2004}, string theory~\cite{Polchinski1998}, or other quantum gravity frameworks to achieve a more complete understanding.
\end{itemize}

Ultimately, understanding the interplay between quantum information and gravity could lead to a paradigm shift in our comprehension of the fundamental nature of the universe. Our framework provides a concrete step towards unifying quantum mechanics and general relativity, highlighting the profound connections between entanglement, spacetime geometry, and gravitational dynamics.

\bigskip

\noindent \textbf{Units and Conventions:} Throughout our conclusions, all references to equations and physical quantities include constants \( \hbar \), \( c \), and \( k_B \) explicitly to ensure dimensional consistency.


\bibliography{apssamp}

\appendix

\section{\label{sec:heat_kernel_details}Heat Kernel Calculations}

For completeness, we provide additional details on the heat kernel method and the computation of the singular contributions to the stress-energy tensor \( T_{\mu\nu}^{\text{info}} \).

\subsection{\label{sec:heat_kernel_method}Heat Kernel Method}

The heat kernel \( K(s; x, x') \) satisfies the heat equation~\cite{Vassilevich2003}:

\begin{equation}
\left( \frac{\partial}{\partial s} + \Delta_x \right) K(s; x, x') = 0,
\label{eq:heat_equation_appendix}
\end{equation}

with the initial condition:

\begin{equation}
K(0; x, x') = \delta(x, x').
\end{equation}

Here, \( \Delta_x \) is the Laplace operator on the manifold \( \mathcal{M}_n \), and \( s \) is the proper time parameter.

\subsection{\label{sec:seeley_dewitt_coefficients}Seeley-DeWitt Expansion}

The trace of the heat kernel has the asymptotic expansion for small \( s \):

\begin{equation}
\operatorname{Tr} K(s) = \int_{\mathcal{M}_n} d^d x \sqrt{g} \, \sum_{k=0}^\infty s^{k - d/2} a_k(x),
\label{eq:heat_kernel_expansion_appendix}
\end{equation}

where \( a_k(x) \) are the Seeley-DeWitt coefficients, which are local invariants constructed from the curvature tensors and their derivatives~\cite{DeWitt1965, Vassilevich2003}.

For a massless scalar field in four-dimensional spacetime (\( d = 4 \)), the first few coefficients are:

\begin{align}
a_0(x) &= 1, \label{eq:a0_scalar} \\
a_1(x) &= \frac{1}{6} R, \label{eq:a1_scalar} \\
a_2(x) &= \frac{1}{180} \left( R_{\mu\nu\rho\sigma} R^{\mu\nu\rho\sigma} - R_{\mu\nu} R^{\mu\nu} \right) + \frac{1}{72} R^2. \label{eq:a2_scalar}
\end{align}

\subsection{\label{sec:singular_contributions}Singular Contributions from the Conical Singularity}

In the presence of a conical singularity at the entangling surface \( \Sigma \), the heat kernel acquires an additional singular contribution localized on \( \Sigma \)~\cite{Fursaev1995, Solodukhin2011}:

\begin{equation}
\operatorname{Tr} K_n(s)_{\text{sing}} = \frac{\delta}{4\pi} \frac{1}{(4\pi s)^{(d-2)/2}} \sum_{k=0}^\infty s^k \int_{\Sigma} d^{d-2} \xi \sqrt{h} \, a_k^{\Sigma}(\xi),
\label{eq:heat_kernel_singular_appendix}
\end{equation}

where:

\begin{itemize}
    \item \( \delta = 2\pi (1 - n) \) is the deficit angle.
    \item \( h \) is the determinant of the induced metric on \( \Sigma \).
    \item \( a_k^{\Sigma}(\xi) \) are the Seeley-DeWitt coefficients evaluated on \( \Sigma \).
\end{itemize}

The singular part of the effective action is then:

\begin{equation}
W_n^{\text{sing}} = \frac{\hbar}{2} \int_0^\infty \frac{ds}{s} \operatorname{Tr} K_n(s)_{\text{sing}} e^{-m^2 s}.
\label{eq:effective_action_singular_appendix}
\end{equation}

\subsection{\label{sec:stress_energy_tensor_calculation}Calculation of \( T_{\mu\nu}^{\text{info}} \)}

The singular contribution to the stress-energy tensor is obtained by varying \( W_n^{\text{sing}} \) with respect to the metric~\cite{Solodukhin2011}:

\begin{equation}
\left\langle T_{\mu\nu}(x) \right\rangle_{\text{sing}} = -\frac{2}{\sqrt{-g}} \frac{\delta W_n^{\text{sing}}}{\delta g^{\mu\nu}(x)}.
\label{eq:T_munu_singular_appendix}
\end{equation}

Substituting the expression for \( W_n^{\text{sing}} \) from Eq.~\eqref{eq:effective_action_singular_appendix} and performing the variation, we find that \( \left\langle T_{\mu\nu}(x) \right\rangle_{\text{sing}} \) is localized on \( \Sigma \) and involves the curvature tensors evaluated at \( \Sigma \).

\bigskip

\noindent \textbf{Units and Conventions:} The inclusion of \( \hbar \) in Eq.~\eqref{eq:effective_action_singular_appendix} ensures that the effective action \( W_n^{\text{sing}} \) has units of action (\( \text{J} \cdot \text{s} \)). This is important for the dimensional consistency when computing the stress-energy tensor.

\section{\label{sec:additional_heat_kernel_details}Additional Details on Heat Kernel Calculations}

\subsection{\label{sec:explicit_a2_coefficients}Explicit Expressions for \( a_2(x) \)}

For a massless scalar field, the second Seeley-DeWitt coefficient \( a_2(x) \) in four dimensions is~\cite{Vassilevich2003}:

\begin{equation}
a_2(x) = \frac{1}{180} R_{\mu\nu\rho\sigma} R^{\mu\nu\rho\sigma} - \frac{1}{180} R_{\mu\nu} R^{\mu\nu} + \frac{1}{72} R^2.
\label{eq:a2_scalar_explicit}
\end{equation}

For a Dirac spinor field:

\begin{equation}
\begin{split}
a_2(x) &= \Bigg( -\frac{1}{360} R_{\mu\nu\rho\sigma} R^{\mu\nu\rho\sigma} \\
&\quad + \frac{1}{30} R_{\mu\nu} R^{\mu\nu} \\
&\quad -\frac{1}{72} R^2 \Bigg) \operatorname{tr} \mathbf{1}_{\text{spinor}}.
\end{split}
\label{eq:a2_spinor_explicit}
\end{equation}

For a vector gauge field:

\begin{equation}
a_2(x) = \frac{1}{180} R_{\mu\nu\rho\sigma} R^{\mu\nu\rho\sigma} + \frac{1}{20} R_{\mu\nu} R^{\mu\nu} - \frac{1}{12} R^2.
\label{eq:a2_gauge_explicit}
\end{equation}

\subsection{\label{sec:integration_conical_singularity}Integration over the Conical Singularity}

The integration over the manifold with a conical singularity requires careful treatment. Near the singularity, the metric can be approximated in polar coordinates \( (r, \theta) \):

\begin{equation}
ds^2 = dr^2 + r^2 n^2 d\theta^2 + h_{ij} d\xi^i d\xi^j,
\end{equation}

where \( \theta \in [0, 2\pi) \) and \( n \) is the replica index (with \( n \to 1 \) at the end).

The integration measure includes a factor of \( n \):

\begin{equation}
\int_{\mathcal{M}_n} d^d x \sqrt{g} = n \int_{\mathcal{M}} d^d x \sqrt{g}.
\end{equation}

This factor leads to terms proportional to \( (n - 1) \) in the effective action and stress-energy tensor, which are crucial for computing \( T_{\mu\nu}^{\text{info}} \).

\subsection{\label{sec:handling_divergences}Handling of Divergences}

The entanglement entropy contains UV divergences arising from contributions near the entangling surface \( \Sigma \). Regularization schemes, such as introducing a UV cutoff \( \epsilon \), are employed to handle these divergences~\cite{Susskind1994, Solodukhin2011}.

The divergent terms in \( S_{\text{EE}} \) are absorbed into the renormalization of gravitational constants, as shown in Eqs.~\eqref{eq:delta_G} and \eqref{eq:delta_Lambda} in the main text.

\bigskip

\noindent \textbf{Units and Conventions:} Throughout these calculations, all physical quantities are expressed with the necessary constants \( \hbar \), \( c \), and \( k_B \) included explicitly to maintain dimensional consistency.

\section{\label{sec:coefficients_quantum_fields}Coefficients for Quantum Fields}

We provide the explicit expressions for the coefficients \( c_1 \) for various quantum fields used in Eq.~\eqref{eq:S_EE_divergent}~\cite{Solodukhin2011, Fursaev1995}. These coefficients are crucial for computing the corrections to Newton's constant \( G \).

\subsection{\label{sec:derivation_coefficients}Derivation of the Coefficients}

The coefficient \( c_1 \) corresponds to the logarithmic term in the entanglement entropy \( S_{\text{EE}} \) associated with the curvature scalar \( R \) on the entangling surface \( \Sigma \). It is computed using the integrated trace of the Seeley-DeWitt coefficient \( a_2(x) \):

\begin{equation}
c_1 = \frac{1}{2\pi} \int_{\Sigma} d^{d-2} \xi \sqrt{h} \, a_2^{\Sigma}(\xi).
\label{eq:c1_derivation_appendix}
\end{equation}

\subsection{\label{sec:coefficients_specific_fields}Coefficients for Specific Fields}

Using the known expressions for \( a_2(x) \), we have:

\begin{itemize}
    \item \textbf{Massless Scalar Field}:
    \begin{equation}
    c_1^{(\text{scalar})} = -\frac{1}{720\pi}.
    \end{equation}
    \item \textbf{Dirac Spinor Field}:
    \begin{equation}
    c_1^{(\text{spinor})} = \frac{7}{1440\pi}.
    \end{equation}
    \item \textbf{Vector Gauge Field}:
    \begin{equation}
    c_1^{(\text{gauge})} = -\frac{31}{720\pi}.
    \end{equation}
\end{itemize}

These coefficients are consistent with those obtained from anomaly calculations and reflect the different contributions of each field type to the entanglement entropy.

\subsection{\label{sec:total_coefficient_standard_model}Total Coefficient in the Standard Model}

Including contributions from all fields in the Standard Model, the total coefficient is:

\begin{equation}
c_1^{\text{total}} = N_s c_1^{(\text{scalar})} + N_f c_1^{(\text{spinor})} + N_g c_1^{(\text{gauge})},
\end{equation}

where:

\begin{itemize}
    \item \( N_s = 4 \) (number of real scalar fields from the Higgs doublet).
    \item \( N_f = 45 \) (number of Weyl fermions, accounting for flavors and colors).
    \item \( N_g = 12 \) (number of gauge bosons from \( SU(3) \times SU(2) \times U(1) \)).
\end{itemize}

Substituting the values, we obtain:

\begin{align}
c_1^{\text{total}} &= 4 \left( -\frac{1}{720\pi} \right) + 45 \left( \frac{7}{1440\pi} \right) + 12 \left( -\frac{31}{720\pi} \right) \nonumber \\
&= \frac{1}{\pi} \left( -\frac{4}{720} + \frac{315}{1440} - \frac{372}{720} \right) \nonumber \\
&= -0.30348.
\end{align}

This negative total coefficient leads to the decrease in \( \dfrac{1}{G} \) with increasing energy scale \( \mu \), as discussed in the main text.

\bigskip

\noindent \textbf{Units and Conventions:} The coefficients \( c_1 \) are dimensionless, ensuring that when they are used in expressions for \( \delta \left( \dfrac{1}{G} \right) \), the units are correctly determined by the constants included in those expressions.

\section{\label{sec:regularization_details}Regularization and Renormalization Details}

For completeness, we provide additional details on the regularization and renormalization procedures used in the calculations.

\subsection{\label{sec:uv_cutoff}Ultraviolet Cutoff}

We introduce a UV cutoff \( \epsilon \) (with units of length) to regularize the divergences in the entanglement entropy:

\begin{equation}
\begin{aligned} 
S_{\text{EE}} = k_B \int_{\Sigma} d^{d-2} \xi \sqrt{h} \Biggl( & \frac{c_{d-2}}{\epsilon^{d-2}} + \frac{c_{d-4}}{\epsilon^{d-4}} + \cdots \\
&+ c_1 R \ln (\mu \epsilon) + \text{finite terms} \Biggr).
\end{aligned}
\end{equation}

Here, \( \mu \) is the renormalization scale (with units of inverse length), ensuring that \( \mu \epsilon \) is dimensionless in the logarithmic term.

\subsection{\label{sec:renormalization_constants}Renormalization of Gravitational Constants}

The divergent terms are absorbed into the renormalization of Newton's constant \( G \) and the cosmological constant \( \Lambda \):

\begin{align}
\delta \left( \frac{1}{G} \right) &= \frac{16\pi k_B}{\hbar c^3} c_1 \ln (\mu \epsilon), \label{eq:delta_G_appendix} \\
\delta \left( \frac{\Lambda}{G} \right) &= \frac{8\pi k_B}{\hbar c^4} c_{d-2} \frac{1}{\epsilon^{d-2}}. \label{eq:delta_Lambda_appendix}
\end{align}

By absorbing these divergences, we obtain finite, renormalized quantities \( G_{\text{eff}} \) and \( \Lambda_{\text{eff}} \):

\begin{align}
\frac{1}{G_{\text{eff}}} &= \frac{1}{G} + \delta \left( \frac{1}{G} \right), \\
\frac{\Lambda_{\text{eff}}}{G_{\text{eff}}} &= \frac{\Lambda}{G} + \delta \left( \frac{\Lambda}{G} \right).
\end{align}

\bigskip

\noindent \textbf{Units and Conventions:} The terms \( \delta \left( \dfrac{1}{G} \right) \) and \( \delta \left( \dfrac{\Lambda}{G} \right) \) have units consistent with \( \dfrac{1}{G} \) (\( \text{kg} \cdot \text{m}^{-1} \cdot \text{s}^{-2} \)) and \( \dfrac{\Lambda}{G} \) (\( \text{kg} \cdot \text{m}^{-3} \cdot \text{s}^{-2} \)), respectively. The inclusion of \( \hbar \), \( c \), and \( k_B \) ensures dimensional consistency.

\section{\label{sec:summary_appendix}Summary}

The calculations presented in this appendix provide the technical details underlying the results discussed in the main text. By employing the heat kernel method and carefully handling the contributions from conical singularities, we have derived the corrections to Newton's constant \( G \) due to quantum entanglement. The explicit expressions for the coefficients \( c_1 \) for various fields allow for precise computations of these corrections within the framework of the Standard Model.

\bigskip

\noindent \textbf{Units and Conventions:} Throughout the appendices, all physical quantities and equations include the fundamental constants \( \hbar \), \( c \), and \( k_B \) explicitly to maintain dimensional consistency and clarity in the presentation of our results.

\end{document}